# Enhanced and Tunable Spin-Orbit Coupling in tetragonally Strained Fe-Co-B Films

R. Salikhov[1*], L. Reichel[2,3], B. Zingsem[1], F. M. Römer[1], R. M. Abrudan[4,5],

J. Rusz[6], O. Eriksson[6], L. Schultz[2,3], S. Fähler[2], M. Farle[1], U. Wiedwald[1]

[1]*Faculty of Physics and Center for Nanointegration (CENIDE), University of Duisburg-Essen, 47057 Duisburg, Germany*

[2]*IFW Dresden, P.O. Box 270116, 01171 Dresden, Germany*

[3]*TU Dresden, Faculty of Mechanical Engineering, Institute of Materials Science, 01062 Dresden, Germany*

[4]*Institute for Condensed Matter Physics, Ruhr-Universität Bochum, 44780 Bochum, Germany*

[5]*Helmholtz-Zentrum-Berlin for Materials and Energy, 12489 Berlin, Germany*

[6]*Uppsala University, Department of Physics and Astronomy, 75120 Uppsala*

E-mail: ruslan.salikhov@uni-due.de

**Abstract**

We have synthesized 20 nm thick films of tetragonally strained interstitial Fe-Co-B alloys epitaxially grown on $Au_{55}Cu_{45}$ buffer layer. The strained axis is perpendicular to the film plane and the corresponding lattice constant *c* is enlarged with respect to the in-plane lattice parameter *a*. By adding the interstitial boron with different concentrations 0, 4, and 10 at.% we were able to stabilize the tetragonal strain in 20 nm Fe-Co films with different *c/a* ratios of 1.013, 1.034 and 1.02, respectively. Using ferromagnetic resonance (FMR) and x-ray magnetic circular dichroism (XMCD) we found that the orbital magnetic moment increases with increasing the *c/a* ratio, pointing towards the enhancement of spin-orbit coupling (SOC) at larger strain. Our results show that careful doping of ferromagnetic films allows to control the SOC by stabilizing anisotropic strain states. These findings are applicable in material design for spintronics applications. We also discuss the influence of B doping on the Fe-Co film microstructure, its magnetic properties and magnetic relaxation.

Spin-orbit interaction is the key phenomenon responsible for coupling of lattice, charge and spin degrees of freedom in a crystal. This fundamentally anisotropic energy allows for the manipulation of magnetization in ferromagnets by electrical current or crystal deformation which is important for the design of spintronic devices [1-3]. The spin-orbit coupling (SOC) directly influences the amplitude of the effective magnetic field arising from the electrical current and acts on the magnetization via spin-orbit torque [2-6]. In order to achieve larger SOC in ferromagnetic (FM) 3d transition metals an interface with heavy (high Z) metals or multilayer layouts are used. Possessing large spin Hall angles, these heavy metal layers act as a "converter" from spin to charge current [7-8] and involve additionally inversion symmetry brake, known as the Rashba effect [4-6]. Interfacing FM layers with nonmagnetic metals brings limitations to the "effective" thickness of the FM layers in use [6, 9-10] and introduces undesirable contributions arising from imperfections like step edges or roughness at the interface. For systematic studies and for designing spintronic devices FM materials with enhanced SOC are of interest, since the conversion of spin current to charge current (or vice versa) is intrinsic in the FM. In that case the spin to charge current conversion efficiency is independent the FM layer's thickness and there is no need for interfacing with additional materials.

The SOC constant $\lambda \sim 0.05$ eV in 3d metals is smaller than the one in 5d transition metals ($\lambda \sim 0.4$ eV). Therefore the phenomena related to SOC, such as spin Hall effect (anomalous Hall effect in case of FM) and magnetic anisotropy energy (MAE) are also expected to be small. It has been shown, however, that the MAE - for example in Fe-Co alloys - can be enhanced by more than three orders of magnitude, if the cubic symmetry of the Fe-Co crystal is broken [11]. This enhancement has been explained by the fact that in the case of small SOC second order perturbation theory can be used in order to account for the MAE. Accordingly, the MAE is inversely related to the energy difference ($\Delta$) between the occupied and unoccupied 3d states. The tetragonal strain in the Fe-Co alloy leads to the reduction of 3d band splitting, $\Delta$, at the Fermi level, which leads to the enhanced MAE [11]. It is noteworthy, that among many transition metals the intrinsic spin Hall conductivity (anomalous Hall conductivity) is significantly larger when the band splitting ($\Delta$) of the d-electrons near the Fermi level is small [2, 12, 13]. These facts suggest that the metastable tetragonal Fe-Co phase can be a very useful material for spintronic applications, since it possesses a large saturation magnetization and possibly a significantly enhanced anomalous Hall conductivity.

Tetragonal phases of Fe-Co alloys can be artificially stabilized by epitaxial growth of ferromagnetic films on a suitable buffer layer [14-18]. Due to the lattice mismatch between the cubic $Fe_{100-x}Co_x$ alloys and a buffer with cubic symmetry the resulting compressive in-plane strain along the axis $a$ leads to a tetragonal distortion with a dilated axis $c$ perpendicular to the film plane. The magnitude of the $c/a$ ratio and the resulting magnetic properties can be tuned by selecting the composition of $Fe_{100-x}Co_x$ or an appropriate buffer layer (see for example [14-18]). An exceptionally large MAE of the order of 1 MJ/m$^3$ recently has been reported for 3 nm Fe-Co film grown on $Au_{50}Cu_{50}$ buffer layer [19]. However, the strain relaxation limits the film thickness with tetragonal distortion to about 4 nm [18]. The large MAE of Fe-Co [19] is the result of both contributions: the tetragonal strain and the hybridization with the $Au_{50}Cu_{50}$ buffer layer. It has been shown recently that B or C doping can be used to stabilize the strain up to a thickness of 100 nm [18, 20-21]. B (or C) preferentially occupies specific octahedral interstitial sites in the $Fe_{100-x}Co_x$ (x < 90) bcc lattice, leading to a local distortion of the position of Fe (or Co) atoms near the doped atom [21] and resulting in a *c/a* ratio of up to 1.045 [19-21]. Measurements of magnetic properties in these interstitial compounds reveal the significantly enhanced MAE of about 0.5 $MJ/m^3$ [21].

The MAE strongly depends on the crystal symmetry and on the filling of the 3*d* band [22] which modifies the energy separation *Δ* of the occupied and unoccupied 3*d* states. The enhanced MAE in Fe-Co-B interstitial compounds [21] could be the consequence of both contributions. We note that the magnetocrystalline anisotropy is proportional to the difference in SOC energies, i.e. the anisotropy of orbital moment of the easy and the hard magnetocrystalline axis [23-24] and it does not directly represent the SOC energies. A direct way to confirm an enhanced SOC is to determine the orbital magnetic moment along the hard magnetocrystalline axis [22]. It can be estimated from the spectroscopic splitting (*g*-) factor, which represents the admixture of orbital moment to the total magnetic moment [23, 25], or by using the sum rules in x-ray magnetic circular dichroism (XMCD) calculating orbital and spin magnetic moments [24, 26-27].

We studied tetragonally strained Fe-Co-B films (0, 4 and 10 at.% B concentration) using ferromagnetic resonance (FMR) and XMCD. The 20 nm thick films were deposited on Au-Cu buffer layers [28]. The thickness of our samples is large enough to neglect the surface/interface contribution to the magnetic properties. Here we show that depending on the B concentration 20 nm Fe-Co films with *c/a* > 1 can be stabilized and that the SOC strength increases noticeably with increasing the *c/a* ratio. The increase of SOC is identified by an increased *g*-factor and out-of-plane (second-order) magnetocrystalline anisotropy parameter $K_2$ – both measured by angular and frequency dependent FMR at room temperature (RT). The ratio of the orbital-to-spin moment ($m_L/m_S$) was estimated by applying the sum rules to the XMCD spectra recorded at the Fe and Co $L_{2,3}$ absorption edges at RT. Our experimental findings suggest the unique possibility of tuning the SOC in ferromagnetic Fe-Co films. This control is independent of the film thickness yielding the possibility to obtain films with a SOC exclusively tuned by B doping without the influence of surface/interface contributions. This makes our tetragonally strained interstitial Fe-Co-B films interesting for spintronic applications. Therefore we also discuss the behavior of the in-plane coercive field and the relaxation rate, the parameters important for designing the elements for spintronics.

The Fe-Co-B (001) samples were prepared using Pulsed Laser Deposition (PLD) in a vacuum chamber with the pressure of 5•10$^{-9}$ mbar at room temperature. Single crystal MgO(100) substrates were used. A 3 nm Cr seed layer and a 30 nm Au$_{55}$Cu$_{45}$ buffer layer were deposited prior the deposition of the 20 nm thick Fe-Co-B films. A 3 nm thick Al capping layer was used to protect samples from oxidation. Using energy dispersive x-ray spectroscopy (EDX) we determined the relative concentrations of Fe and Co to be 40:60 (Fe$_{40}$Co$_{60}$) for all samples. The boron content was equivalent to the compositions from Ref. [21] 0, 4 and 10 at.% B. The final composition of the ferromagnetic layers was thus (Fe$_{40}$Co$_{60}$)$_{1-y}$B$_y$ with the y = 0, 0.04 and 0.1 for three different samples. A Bruker D8 Advance diffractometer operating with Co $K_\alpha$ radiation was used for x-ray diffraction (XRD) in Bragg-Brentano geometry. The Fe-Co-B lattice strain (*c/a* ratio) in our samples was determined from Fe-Co(011) pole figure measurements using an X'pert four circle goniometer setup with Cu $K_\alpha$ radiation. For a more detailed description of our sample preparation and structural characterization we refer to [21].

Magnetic hysteresis measurements were performed using a vibrational sample magnetometer (VSM). In order to determine the *g*-factors two FMR setups were used. At first, the FMR measurements were performed using a conventional Bruker X-band FMR spectrometer operated at a microwave (mw) frequency of $f$ = 9.78 GHz. The FMR spectra were recorded by sweeping the external magnetic field $H$ at different angles $\varphi_H$ with respect Fe-Co (110) crystallographic direction within the film plane. Secondly, we performed FMR measurements using a semi-rigid mw cable short-circuited at its end. This setup allows for field dependent FMR measurements at different mw frequencies, ranging from $f$ = 2 – 26 GHz [29]. The measurements were performed by sweeping an external magnetic field applied parallel to Fe-Co [110] in-plane magnetic easy axis at fixed frequencies. Both, in-plane angular dependence and the frequency dependent measurements were fitted simultaneously with identical set of parametres. Each FMR spectrum was fitted using a general analytic solution of the full tensor representation of the high frequency susceptibility [30] derived from the Landau-Lifshitz-Gilbert equation [23]. In order to fit the complete FMR line shape the fitting model includes an asymmetric excitation to account for the asymmetric resonance lines. For the in-plane angular ($\varphi_H$) dependence of the resonance magnetic field ($H_r$) we solved the Smit-Beljers equation [31] in the respective equilibrium states of the magnetization as described in [32]. For all fits we use the free energy density given in Eq. (1), which includes the demagnetizing energy term, the anisotropic energy density for the tetragonal symmetry and the Zeeman energy density:

$$F = (\frac{1}{2}\mu_0(N_\perp - N_\parallel)M_s^2 - K_2)\cos^2\theta - \frac{1}{8}K_4^\parallel(3 + \cos 4\varphi)\sin^4\theta - \mu_0 M_s H\,(\cos\theta\cos\theta_H$$
$$+ \cos(\varphi - \varphi_H)\sin\theta\sin\theta_H)\,, (1)$$

with $(\theta, \varphi)$ and $(\theta_H, \varphi_H)$ being the polar (accounted from the film plane normal) and azimuthal (in-plane) angles of the magnetization $M_s$ and the applied magnetic field $H$, respectively. We use the demagnetization factors $N_\perp = 1$, $N_\parallel = 0$ of an infinitely thin film, which is appropriate for the respective thickness of our films [33]. $K_2$ and $K_4^\parallel$ are the second- and fourth-order terms of the MAE density [23]. The saturation magnetization $M_s$ was determined from VSM measurements, thus, the $g$-factor, the damping parameter $\alpha$, the out-of-plane magnetocrystalline anisotropy second-order constant $K_2$ and the in-plane forth-order magnetic anisotropy constant $K_4^\parallel$ were used as fit parametres.

The x-ray absorption spectra (XAS) were recorded in total electron yield (TEY) using the ALICE chamber [34] as an end station at the PM3 beamline at the synchrotron radiation facility BESSY II of the Helmholtz Zentrum Berlin. The beamline provides intense x-ray beams at the photon energies of the Fe and Co $L_{3,2}$ absorption edges with 95% of circular polarization. XAS was recorded at the Fe and Co $L_{3,2}$ absorption edges. Spin and orbital contributions of Fe and Co magnetic moment were calculated form corresponding XMCD spectra using the sum rules [24, 26-27]. Measurements were performed at different angles of incidence (20º - 90º) with respect to the sample surface in order to correct for saturation effect [27, 35]. The wave vector of the incoming beam was parallel to the in-plane Fe-Co <110> magnetic easy axis for probing the magnetic moment. Element-specific hysteresis loops for all samples show a rectangular shape with 100% remanence. Prior to starting the energy scans our samples were magnetically saturated at $\mu_0 H = 0.27$ T applied parallel to the Fe-Co (110) direction. Energy scans were made in zero fields. Both FMR and XMCD measurements were performed at the room temperature.

XRD and pole figure measurements confirm the epitaxial growth of the Fe-Co-B films with (001) growth direction on the Au-Cu buffers. The only XRD peaks of the Fe-Co-B films in the Bragg-Brentano scans correspond to Fe-Co (002) and are shifted to lower angles with respect to the one expected for bcc $Fe_{40}Co_{60}$. This shift points towards an increased lattice parameter along the out-of-plane direction (*c*-axis) with respect to the lattice parameter of bcc $Fe_{40}Co_{60}$. With increase of B content the intensity of the Fe-Co (002) reflections decreases and peaks broaden, indicating that B doping strongly supports the reduction of film crystallinity. In order to quantify this relative decrease of the crystal size we have calculated the x-ray coherence length using Scherrer's formula [36]. The calculated values are plotted in Fig. 1(a). The tetragonal strain was characterized from {011} pole figure measurements as described in Ref. [18, 21]. The calculated *c/a* ratios for all three samples are presented in Fig. 1(b) and Table I. An initial strain with *c/a* = 1.013 is observed in the binary $Fe_{40}Co_{60}$ reference sample and might be attributed to a preparation related strain in film growth direction, which typically has an order of 1% [37-38]. However, one can see that 4 at.% B supports the tetragonal strain stabilization with significantly enhanced *c/a* ratio of 1.034. Subsequent increase of B content up to 10 at.% leads to a decrease of tetragonal strain, but the *c/a* ratio of 1.02 remains larger than the one in the binary Fe-Co film. The decrease of the strain at larger B content is the result of limited solubility of B to form the Fe-Co-B interstitial compounds. At 10 at.% this limit is reached and B tends to form B-rich phases with less distortion instead of occupying the Fe-Co interstitials along the *c*-axis as has been described in Ref. [21] in detail. We conclude that the largest distortion can be achieved only for a carefully selected B concentration smaller than 10 at.%.

Next we show that the reduced crystallite dimensions result in a softening of in-plane magnetic properties. The in-plane angular dependence of the FMR field ($H_r$) with respect to Fe-Co (110) direction for all systems is shown in Fig. 2(a). The solid lines represent the fit to the experimental data points. From the graph it is apparent that for all samples the magnetic easy axis is lying in the film plane, since $\mu_0 H_r < 2\pi\mu_0 f/\gamma \approx 300$ mT ($\gamma = g\mu_B/\hbar$ is the absolute value of the electron gyromagnetic ratio) [23]. One can also see that all ferromagnetic films exhibit four-fold symmetry with the magnetocrystalline easy axis parallel to Fe-Co (110). The values of $K_4^{\parallel}$ obtained from the fit are plotted in Fig. 1(a) together with the values of the coercive field ($H_c$) determined from the hysteresis loops along the Fe-Co (110) using the ALICE [34]. Both $K_4^{\parallel}$ and $H_c$ are decreasing with increasing B content, showing excellent correlation with the decrease of the XRD coherence length (Fig. 1(a)). This confirms that the reduction of magnetocrystalline anisotropy scales with the crystallinity of our epitaxial films and causes the magnetic softening with noticeably smaller coercive fields (Fig. 1(a)). In the following we show, that despite the fact that alloying with B leads to a softening of in-plane magnetic properties the SOC in Fe-Co-B samples increases.

Fig. 2(b) presents examples of FMR spectra recorded at different frequencies including their respective fit. The thickness of Fe-Co ferromagnetic layers was selected such that contribution of eddy currents and skin depth is negligible [39-40]. All FMR spectra measured at different frequencies were fitted simultaneously taking into account an effective asymmetry in the projection between excitation and measurement of the high frequency susceptibility [30] and consistent with the angular dependent spectra. The values of $K_2$ and the *g*-factor together with the *c/a* ratio are plotted in Fig. 1(b). The *g*-factor follows the *c/a* ratio: it reaches a maximum of 2.18 at 4 at.% B, has an intermediate value of 2.14 at 10 at.% B and is minimal, 2.11, for the reference Fe-Co film. $K_2$ shows a different trend, like the g-factor it reaches a maximum at 4 at.% B concentration, but at 10 at.% it nearly vanishes. The fact that it is almost zero at the smallest crystal grain size at 10 at.% B concentration indicates that the macroscopic MAE is not a good indicator for the SOC in that alloy film.

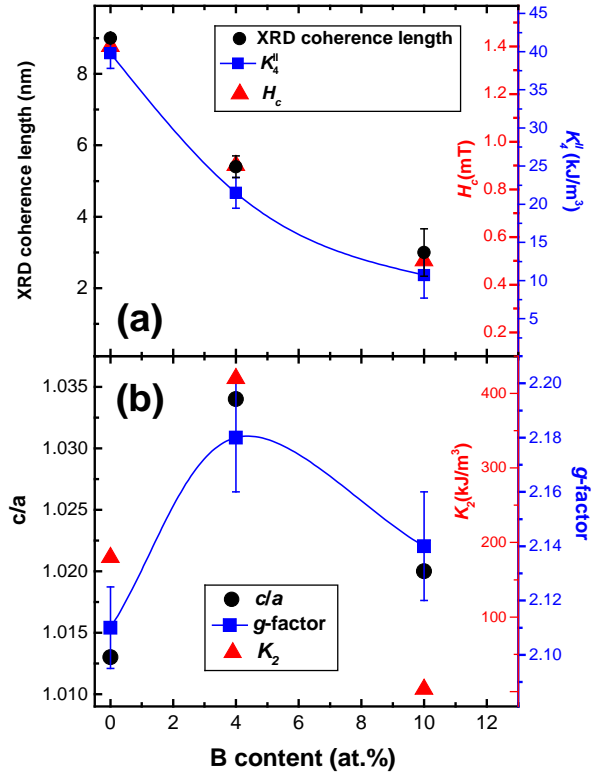

Fig. 1. Dependence of structure and magnetism on the B concentration in Fe-Co films: (a) XRD coherence length (black circles and left scale bar), in-plane 4-fold MAE density $K_4^{\parallel}$ (blue squares and right outer scale bar) and coercive field $H_c$ (red triangles and right inner scale bar); (b) c/a ratio (black circles and left scale bar), g-factor (blue squares and right outer scale bar) and out-of-plane 2-fold MAE density $K_2$ (red triangles and right inner scale bar). Solid lines are guide to the eye.

The g-factor in *3d* metals can be estimated similar as for noninteracting electrons [23, 37-38] using the equation (2) for small SOC in first order approximation

$$g = 2.0023 \left(\frac{m_L}{m_S} + 1\right). \quad (2)$$

In absence of SOC, the orbital moment is completely quenched and the g-factor is similar to the one for free electrons (g = 2.0023) [23, 41-42]. An increased SOC shifts the g-factor to higher values. Therefore, one can conclude that the SOC in our samples scales with the *c/a* ratio and, thus, can be intentionally adjusted by choosing the B concentration. The g-factor in the binary reference sample is larger than the one reported for epitaxial Fe-Co films with identical composition [43]. This indicates that the PLD related tetragonal strain is already contributing to the SOC. The relative increase in the *g*-factor (SOC) with 4 at.% B doping is significant and is comparable to, e.g. the increase in the *g*-factor in 1 nm thick Py film due to sandwiching from both sides by Pt or Ta [44]. For the further direct comparison with the XMCD results we have calculated the values of $\frac{m_L}{m_S}$ ratio using Eq. (2) (Table I.)

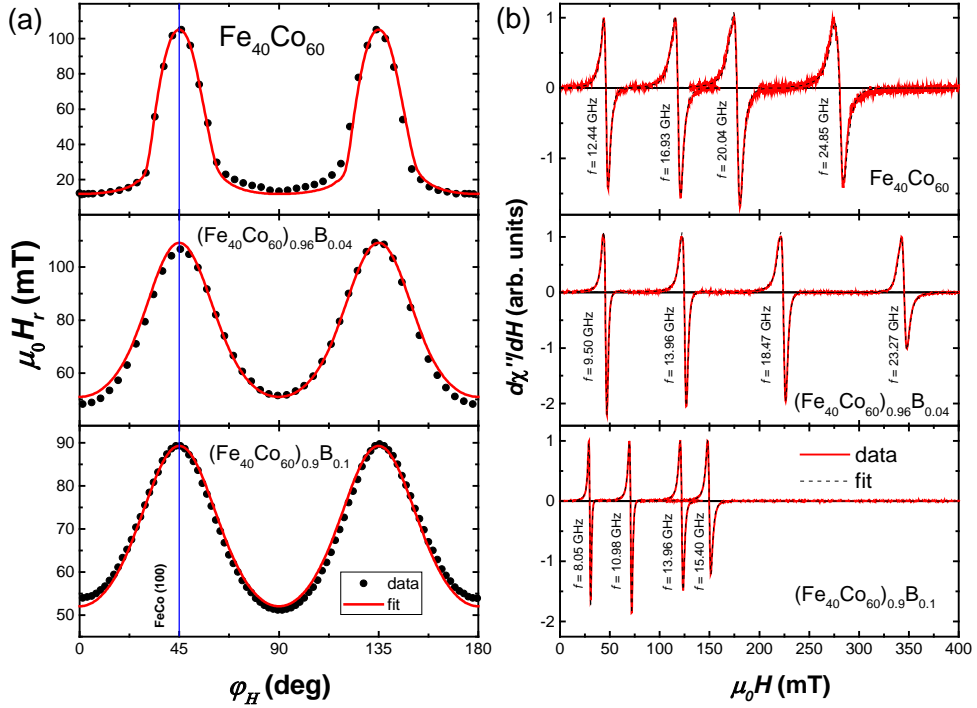

Fig. 2. (a) In-plane angular dependence of resonance field $H_r$ measured by FMR at $f$ = 9.78 GHz for studied Fe-Co films with different B concentration. Angles $\varphi_H$ are counted with respect to $Fe_{40}Co_{60}$ [110] direction. The red solid lines represent the fit to the data points (black circles). The slight mismatch between the fit and experimental data is the result of presence a small in-plane uniaxial anisotropy which does not influence the fit parameters significantly. (b) Some examples of FMR spectra (red solid lines) recorded at different frequencies using semi rigid short-circuited mw cable for three studied samples. Dashed black lines in all spectra represent the fit.

XAS spectra were detected by TEY. The probing depth of TEY is comparable to the effective escape depth of secondary electrons which is about 2 nm [27, 35], therefore the interface between the 20 nm thick ferromagnetic films and Au-Cu buffer does not influence the detected XAS and XMCD spectra. XAS measured at Fe $L_{3,2}$ absorption edges at two different x-ray beam helicities for the $Fe_{40}Co_{60}$ and $(Fe_{40}Co_{60})_{0.96}B_{0.04}$ are shown in Fig. 3 (a) and (b) respectively. A pure metallic character of the spectra is seen for all samples. The integral absorption intensity $[I(L_2) + I(L_3)]$ over $L_3$ and $L_2$ edges of the "white-line" spectra is similar for all studied samples. However, the branching ratio $I(L_3)/[I(L_2) + I(L_3)]$ changes significantly between the spectra shown in Fig. 3 (a) and (b). This indicates that the splitting of 3*d* electronic levels due to spin-orbit interaction is different for those two magnetic systems [45]. XAS recorded at Co-edges shows identical behavior. In order to quantify the change in SOC we calculated the $\frac{m_L}{m_S}$ ratio using the sum rules. XMCD curves for Fe$_{40}$Co$_{60}$ and (Fe$_{40}$Co$_{60}$)$_{0.96}$B$_{0.04}$ films derived from XAS at the Fe and Co edges are shown in Fig. 3 (c) and (d) respectively. Magnetic asymmetry for both Fe and Co atoms reduces with B doping. This is in agreement with VSM measurements where we find $\mu_0 M_s$ to be 2.1 T, 1.8 T and 1.4 T for samples with 0, 4 and 10 at.% B respectively. First we discuss the results for Fe$_{40}$Co$_{60}$ reference sample. We obtain following spin ($m_S$) and orbital ($m_L$) magnetic moments from XMCD of Fe and Co spectra:

Fe: $m_S^{Fe} = 2.2$ μ$_B$/atom; $m_L^{Fe} = 0.1$ μ$_B$/atom

Co: $m_S^{Co} = 1.73$ μ$_B$/atom; $m_L^{Co} = 0.15$ μ$_B$/atom.

Comparison of these results with the values calculated using the relativistic scattering theory in Ref. [46] for the Fe-Co alloy with corresponding concentrations reveals that both the Fe and Co spin moments are underestimated by about 15%. We also found that orbital moments from the experiment are larger than the values in Ref. [46]. The relatively small disagreement in spin magnetic moments can be the result of the contribution of intra-atomic magnetic dipole correction term $T_Z$, which we assume to be significant for our strained films [47]. The tetragonal strain in our sample could also be the reason of enhanced orbital moments with respect to bcc Fe$_{40}$Co$_{60}$. Based on our XMCD data we have calculated the averaged contribution of spin and orbital moment per magnetic atom for the Fe$_{40}$Co$_{60}$ film. This allows us to estimate the value of the magnetic polarization and the *g*-factor (using Eq. (2)). We found the magnetic polarization to be about 2.08 T and $g = 2.13$. Considering the experimental errors and uncertainness given by $T_Z$ term both parameters are in a good agreement with VSM ($\mu_0 M_s = 2.1$ T) and FMR ($g = 2.11 \pm 0.015$) measurements. Beside the $T_Z$ term for samples (Fe$_{40}$Co$_{60}$)$_{0.96}$B$_{0.04}$ and (Fe$_{40}$Co$_{60}$)$_{0.9}$B$_{0.1}$ with larger strain, boron doping can also introduce uncertainty for the application of the sum rules modifying the polarization of 4*s* electrons or, in another words, diffuse magnetic moments [27]. The modification of the diffuse magnetic moments could explain the disappearance of small positive intensity shoulder right above the XMCD Fe-$L_3$ peak for (Fe$_{40}$Co$_{60}$)$_{0.96}$B$_{0.04}$ sample when comparing with Fe$_{40}$Co$_{60}$ in Fig. 3(c). These features certainly contribute to the integrals for calculating spin and orbital moments. Nevertheless, as observed from modification of the branching ratio in "white-line" spectra the change of orbital contribution with B doping is significant. Therefore the relative change of the orbital moment determined from the sum rules within the set of studied samples can be compared. In order to avoid the error, which arises due to the influence of B doping (which also leads to visible decrease of Fe-Co magnetic moment) on the 3*d* electron occupation number (number of valence *d* holes) we have calculated the relative ratio between orbital and spin magnetic moments [26]. Effective $\frac{m_L}{m_S}$ ratios for all studied samples are presented in Table I together with the ratios determined from FMR studies.

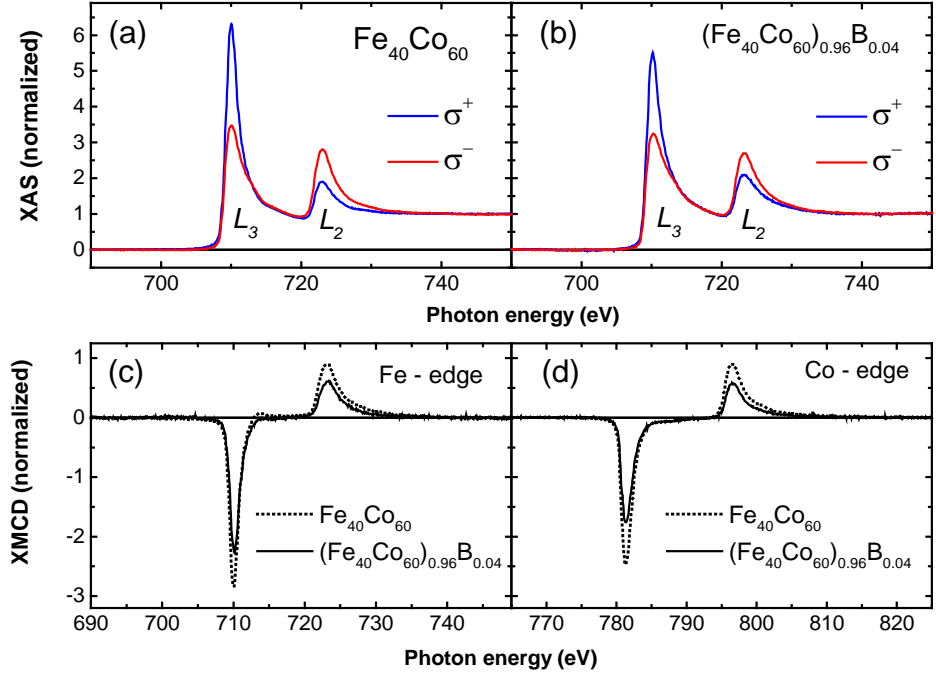

Fig. 3. Top panel: Normalized Fe $L_{3,2}$-edge XAS detected by TEY for right ($\sigma^+$) and left ($\sigma^-$) circularly polarized x-rays of $Fe_{40}Co_{60}$ (a) and $(Fe_{40}Co_{60})_{0.96}B_{0.04}$ (b) films. Bottom panel: Comparison of normalized XMCD spectra for $Fe_{40}Co_{60}$ (dashed line) and $(Fe_{40}Co_{60})_{0.96}B_{0.04}$ (solid line) measured at the Fe-edge (c) and Co-edge (d).

TABLE I. $\frac{m_L}{m_S}$ ratio determined from XMCD and FMR measurements for tetragonally strained Fe-Co-B films with different *c/a* ratio.

| Sample | *c/a* ratio | $\frac{m_L}{m_S}$ (FMR) | $\frac{m_L}{m_S}$ (Fe-edge) | $\frac{m_L}{m_S}$ (Co-edge) |
|---|---|---|---|---|
| $Fe_{40}Co_{60}$ | 1.013 | 0.055 | 0.04 | 0.085 |
| $(Fe_{40}Co_{60})_{0.96}B_{0.04}$ | 1.034 | 0.09 | 0.1 | 0.14 |
| $(Fe_{40}Co_{60})_{0.9}B_{0.1}$ | 1.02 | 0.07 | 0.06 | 0.14 |

It is evident, that the stabilization of the tetragonal strain with larger *c/a* ratio by B doping leads to a relative increase in orbital magnetic moment of both Fe and Co atoms. The $\frac{m_L}{m_S}$ ratio determined from the spectra recorded at the Fe absorption edges follows the trend given by the *c/a* ratio and $\frac{m_L}{m_S}$ from the FMR measurements: the larger the lattice strain the bigger the orbital admixture to the magnetic moments of the Fe atoms. However, for 4 at.% B sample the $\frac{m_L}{m_S}$ ratios of Fe and Co are larger than the ones estimated from FMR. This is most likely the result of an additional contribution from the intra-atomic dipole moment to the spin magnetic moments in this highly strained sample [47]. In contrast to Fe the calculated $\frac{m_L}{m_S}$ ratio for Co shows identical values for 4 and 10 at.% B samples. We attribute this to the significantly increased errors using the sum rules for the Fe-Co film with 10 at.% B.

Spin-orbit interaction is also responsible for magnetic relaxation processes in ferromagnetic systems, since this is the general mechanism of coupling the spin degree of freedom to the lattice (phonons) [33, 39]. The narrowing of the FMR line with B doping is clearly seen in Fig. 2(b), which indicates that the magnetic relaxation rate decreases with increasing B concentration. In order to compare the intrinsic contributions to the overall relaxation rate in our samples we plot the values of FMR peak-to-peak linewidth ($\Delta H_{pp}$) as a function of mw frequency $f$ in Fig. 4. For the highly doped $(Fe_{40}Co_{60})_{0.9}B_{0.1}$ system the FMR linewidth (not shown) doesn't exhibit a clear linear dependence on the excitation frequency pointing towards a significant contribution of two-magnon scattering to the magnetic relaxations [48] in this film with smallest crystallinity. The solid red lines in Fig. 4 represent the fit to the data from $Fe_{40}Co_{60}$ and $(Fe_{40}Co_{60})_{0.96}B_{0.04}$ films using equation (3).

$$\Delta H_{pp} = \Delta H_0 + \frac{2}{\sqrt{3}}\frac{\alpha}{\gamma}2\pi f, (3)$$

where $\Delta H_0$ is the inhomogeneous broadening of the FMR linewidth. The inhomogeneous (extrinsic) broadening for both samples is low and does not exceed $\mu_0 \Delta H_0 = 1.5$ mT. It increases slightly with 4 at.% B doping. From the fit we find that the intrinsic damping parameter $\alpha$ (presented in Fig.4) for the $(Fe_{40}Co_{60})_{0.96}B_{0.04}$ system is smaller than in binary $Fe_{40}Co_{60}$. This indicates that the significant increase of SOC in 4 at.% B doped sample doesn't lead to an increase in the intrinsic relaxation rate. This behavior is possible in systems with a long spin flip relaxation time ($\tau_{sf}$), where relaxation can be described by the "breathing Fermi surface" mechanism [49], which is valid for both pure Fe and pure Co films [39]. In that particular case the damping parameter is proportional to electrical conductivity, therefore a decrease in $\alpha$ in the 4 at.% B film could be explained by reduced conductivity due to an almost two times smaller crystallinity found from XRD measurements (Fig.1 (a)). The decrease in $\alpha$ for the sample with large SOC indicates that in this film $\tau_{sf}$ is still large, at least two orders of magnitude larger than for Py films [39].

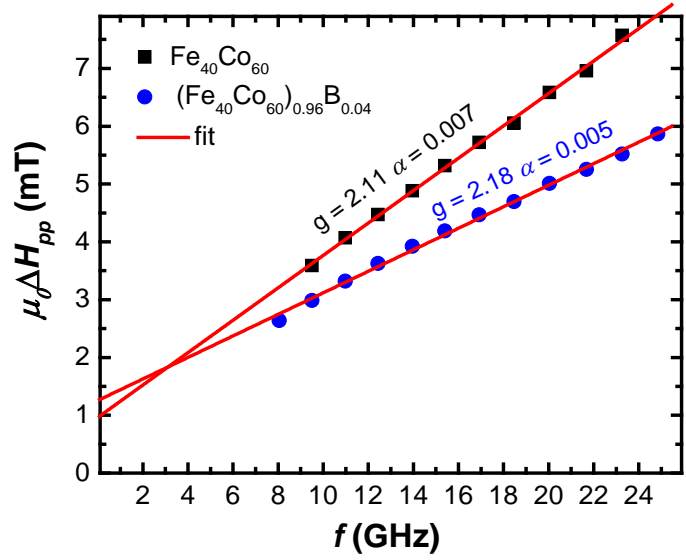

Fig. 4. The frequency dependence of peak-to-peak FMR linewidth $\Delta H_{pp}$ determined from the fit to the spectra after correcting for the asymmetry of the absorption line for $Fe_{40}Co_{60}$ (black squares) and $(Fe_{40}Co_{60})_{0.96}B_{0.04}$ (blue circles) systems. The red solid lines represent the fit using equation (3).

In conclusion using PLD we have synthesized 20 nm thick tetragonally distorted Fe-Co-B films with $c/a$ ratios up to 1.034. The ratio of orbital-to-spin magnetic moment increases with increasing $c/a$ ratio, representing the increased admixture of orbital magnetic moment to the total magnetic moment. This indicates that SOC energy scales with the $c/a$ ratio in our systems. Our results offer an approach to control spin-orbit interaction in ferromagnetic films for spintronic applications. B doping leads to a reduced crystallinity and magnetic softening of the in-plane hysteresis loop with a coercivity comparable to the one known for permalloy. Focusing our attention on the magnetic relaxation characteristics we find that the intrinsic damping parameter $\alpha$ decreases in samples with larger SOC. This fact can only be explained assuming a long spin flip relaxation time of conduction electrons in all studied samples.

### Acknowledgements

We acknowledge funding from the European Community´s Seventh Framework Programme (FP7-NMP) under grant agreement no. 280670 (REFREEPERMAG). We are also thankful to the Helmholtz Zentrum Berlin for travel support.